\documentclass[twocolumn,showpacs,preprintnumbers,amsmath,amssymb]{revtex4}

\usepackage{graphicx}
\usepackage{dcolumn}
\usepackage{bm}

\begin{document}

\def\la{\mathrel{\mathpalette\fun <}}
\def\ga{\mathrel{\mathpalette\fun >}}
\def\fun#1#2{\lower3.6pt\vbox{\baselineskip0pt\lineskip.9pt
\ialign{$\mathsurround=0pt#1\hfil##\hfil$\crcr#2\crcr\sim\crcr}}} 


\title{Rapidity-dependence of jet shape broadening and quenching}

\author{I.P.~Lokhtin}
\author{S.V.~Petrushanko} 
\author{L.I.~Sarycheva}
\author{A.M.~Snigirev}
\affiliation{ M.V.~Lomonosov Moscow State University, D.V.~Skobeltsyn Institute of Nuclear Physics, 
119992,  Leninskie Gory, Moscow, Russia}

\date{\today}

\begin{abstract}
The jet shape modification due to partonic energy loss in the dense QCD matter is
investigated by the help of the special transverse energy-energy correlator
in the vicinity of maximum energy deposition of every event.
In the accepted scenario with scattering of jet hard partons off 
comoving medium constituents this correlator is independent of the pseudorapidity 
position of a jet axis and becomes considerably broader (symmetrically over the 
pseudorapidity and the azimuthal angle) in comparison with $pp$-collisions.
At scattering off  ``slow''  medium constituents the broadening of  
correlation functions is dependent on the pseudorapidity position of a jet axis 
and increases  noticeably in comparison with the previous scenario for jets with 
large enough pseudorapidities. These two considered scenarios result also in the
different dependence of jet quenching on the pseudorapidity.
\end{abstract}

\maketitle

\section{INTRODUCTION}
Experimental results from the CERN Super Proton Synchrotron (SPS)
have revealed a rich set of new
phenomena as the anomalous suppression of charmonium~\cite{alex} and the
broadening or mass shift of vector mesons~\cite{ceres} indicating the formation
of dense matter~\cite{heinz, satz}. However, it has proven difficult to
disentangle hadron contributions to the observed signals, and no clear
consensus has emerged that a long lived quark-gluon plasma (QGP)
has been observed in
the fixed target heavy ion experiments at those facilities. Experiments at
the BNL Relativistic Heavy Ion Collider (RHIC), 
which began operation in 2000, have initiated a new era in the study of
QCD matter under extreme conditions. Nuclear collisions at the largest RHIC
energy ($\sqrt{s_{NN}}=200$ GeV) not only  produce matter at the highest energy
density ever achieved but also provide a number of rare observables that have
not been accessible previously. Already in the first three years of RHIC 
operation the wealth of
new data has been collected and analyzed such as a strong elliptic flow, the
suppression of hadron spectra and azimuthal back-to-back two-particle
correlations indicating that a dense equilibrated system is generated in the
most violent head-on collisions of heavy nuclei (see, for instance,
a review~\cite{jacob} and references therein).

The c.m. energy for heavy ion collisions at the CERN Large Hadron Collider (LHC) 
($\sqrt{s_{NN}}=5.5$ TeV) will exceed that at RHIC by a factor of about 30.
This provides exciting opportunities for  addressing unique physics issues in a
new energy domain, where hard and semi-hard  QCD multi-particle production can
stand out against the underlying soft events. Besides the statistics is
expected to be high enough~\cite{Accardi:2003} for the  systematic 
analysis of various aspects of
QCD-physics (poorly accessible for studying at RHIC energy) in the medium with the
initial energy density considerably above the critical one for the
deconfinement transition. In particular it will be  possible to study the
modification of jet shape and spectra in comparison with $pp$-collisions.
Jets as one of the important probes of quark-gluon plasma  are created at
the very beginning of the collisions process ($\tau_{{\rm form}} \sim
1/p_T \la 0.01$ fm/$c$) by the initial hard parton-parton scatterings. These hard
partons pass through the dense matter formed at longer time scales
($\ga 0.1$ fm/$c$) and interact strongly with the medium constituents changing
its original properties (the momentum direction, energy, particle distribution
inside a jet) as a result of in-medium rescattering. 

One of the encouraging
methods for investigation of these effects seems to be the measurement of
energy-energy correlations~\cite{pan94, lokhtin04}. In the present paper we
extend our approach~\cite{lokhtin04}, considering the energy-energy correlator
in the restricted rapidity-angle space in the vicinity of maximum energy
deposition of every event. This special correlator allows us to investigate,
using the calorimetric information,
the broadening of the jet shape due to partonic energy loss which is
intensively discussed in the current literature~\cite{urs04}.

\section{ENERGY CORRELATORS IN ELECTRON-POSITRON,
HADRON AND NUCLEAR COLLISIONS}
To be clear at first we recall the well-known definition of the energy-energy
correlation function $\Sigma$  which has been used by all for LEP
experiments~\cite{abreu} at CERN and the SLD experiment~\cite{abe} at SLAC to 
measure the strong coupling constant $\alpha_s$ in $e^+e^-$-annihilation at 
the $Z^0$ resonance with a high accuracy.
$\Sigma$ is defined as a function of the angle $\chi$ between two particles $i$
and $j$ in the following form: 
\begin{equation}
\label{1}
\frac{d\Sigma(\chi)}{d\cos(\chi)} = \frac{\sigma}{\Delta \cos(\chi) N_{\rm event}}
\sum \limits_{{\rm event}} \sum\limits_{i\not= j} \frac{E_iE_j}{E^2}, 
\end{equation}
where $\sigma$ is the total cross section for $e^+e^- \to$ hadrons,
$E$ is the total energy of the event, $E_i$ and $E_j$ are the energies of
the particles $i$ and $j$. The sum runs over all pairs $i$, $j$ with
$\cos(\chi_{ij})$ in a bin width $\Delta \cos(\chi)$: 
$$\cos(\chi) - \Delta\cos(\chi)/2 < \cos(\chi_{ij}) < \cos(\chi) +
\Delta\cos(\chi)/2.$$
The limits
$\Delta\cos(\chi) \to 0$ and $N_{\rm event} \to \infty$ have to be taken in
Eq. (\ref{1}).

This function can be calculated in perturbative QCD as a series in $\alpha_s$:
\begin{eqnarray}
\label{2}
\frac{1}{\sigma_0} \frac{d\Sigma(\chi)}{d\cos(\chi)} =&&
\frac{\alpha_s(\mu)}{2\pi}A(\chi) 
+\left(\frac{\alpha_s(\mu)}{2\pi}\right)^2 \\
&&\times\left(\beta_0 \ln\left(\frac{\mu}{E}\right) A(\chi)+B(\chi)\right) +
O(\alpha_s^3),\nonumber
\end{eqnarray}
where $\beta_0 = (33-2n_f)/3$, $n_f$ is the number of active flavors at the
energy $E$, $\sigma_0$ is the Born cross section, $\mu$ is the renormalization
scale.  
The first order term $A(\chi)$ has been calculated by Basham {\it et al.}
\cite{basham} from the well-known one gluon emission diagrams $\gamma^*,Z^0 \to q\bar
qg$ with the result 
\begin{eqnarray}
\label{3}
A(\chi) = &&C_F(1+\omega)^3\frac{1 + 3\omega}{4\omega} ((2-6\omega^2)\nonumber\\ 
&&\times\ln(1 +1/\omega) + 6\omega - 3),
\end{eqnarray}
where $C_F=4/3,~\omega = \cot^2(\chi)/2$ and 
$\chi$ is the angle between any of the three partons initiating the three 
hadron jets. 
This allowed one to determine the strong coupling constant directly
from a fit to the energy-energy correlations in $e^+e^-$-annihilation, since
$\Sigma$ is proportional to $\alpha_s$ in the first order. 

In hadronic and nuclear collisions jets are produced by hard scatterings of
partons. In this case it is convenient to introduce the transverse energy-energy
correlations which depend very weakly on the structure functions \cite{ali} and
manifest directly the topology of events. Thus, for instance, in high-$p_T$
two-jet events the correlation function is peaked at the azimuthal angle $\varphi =
0^\circ$ and $\varphi = 180^\circ$, while for the 
isotropic background this function is independent of $\varphi$. 

Utilization of hard jet characteristics to investigate QGP in heavy ion 
collisions is complicated because of a huge multiplicity of ``thermal'' 
secondary particles in an event. Various estimations give from 1500 to 6000 
charged particles per rapidity unit in a central Pb$+$Pb collision at LHC energy, 
and jets can be really reconstructed against the background of energy flux 
beginning from some threshold jet transverse energy $E_T^{\rm jet} \sim 50-100$ 
GeV \cite{Accardi:2003, kruglov}. The transverse energy-energy correlation function 
is just sensitive to the jet quenching as well as to the global structure of 
energy flux (i.e. its anisotropy for non-central collisions) if we select 
events by an appropriate way~\cite{lokhtin04}. The generalization of 
Eq. (\ref{1}) for calorimetric  
measurements of the energy flow is straightforward: 
\begin{equation}
\label{4}
\frac{d\Sigma_T(\varphi)}{d\varphi} = \frac{1}{\Delta\varphi~ N_{\rm
event}}\sum\limits_{{\rm event}}\sum\limits_i \frac{E_{Ti}
E_{T(i+n)}}{(E_T^{\rm vis})^2},
\end{equation}
where
$$E_T^{\rm vis} = \sum\limits_i E_{Ti}$$
is the total transverse energy in $N$ calorimetric sectors covering the full 
azimuth, $E_{Ti}$ is the transverse energy deposition in the sector $i$
($i=1,...,N$) in the 
considered pseudorapidity region $|\eta|$, $n=[\varphi/\Delta\varphi]$ (the 
integer part of the number $\varphi/\Delta\varphi$), $\Delta\varphi \sim 0.1$
is the typical azimuthal size of a calorimetric sector. In the continuous 
limit $\Delta\varphi \to 0$, $N = [2\pi/\Delta\varphi] \to \infty$ 
Eq. (\ref{4}) reads 
\begin{equation}
\label{Sigma_T}
\frac{d\Sigma_T(\varphi)}{d\varphi} = \frac{1}{N_{\rm
event}}\sum_{{\rm event}}\frac{1}{(E_T^{\rm vis})^2}\int\limits_0^{2\pi}d\phi
\frac{dE_T(\phi)}{d\phi}\frac{dE_T(\phi + \varphi)}{d\phi},
\end{equation}
where
$\displaystyle E_T^{\rm vis} = \int\limits_0^{2\pi} 
d\phi~\frac{dE_T(\phi)}{d\phi}$, 
~~$\displaystyle \frac{dE_T}{d\phi}$ is the distribution of transverse energy flow
over azimuthal
angle $\phi$. 

In our previous work~\cite{lokhtin04} it has been shown that
at the special selection of events for the analysis (at least one
high-$p_T$ jet) and the procedure of event-by-event background subtraction 
in high multiplicity heavy ion collisions, the transverse energy-energy 
correlator (\ref{4})  is sensitive to the partonic energy loss and angular spectrum of 
radiated gluons. The medium-modified energy-energy correlation
function manifests significant strengthening in a wide interval of azimuthal 
angles around $\pi/2$, moderate broadening of the near-side jet region  
$\varphi \la 0.5$ and weak additional suppression of back-to-back correlations 
for $\varphi \sim \pi$. Without jet trigger 
this correlation function shows the global structure of transverse energy flux:
the correlator is isotropic for central collisions and for non-central 
collisions it is sensitive to the azimuthal anisotropy of energy flow 
reproducing its Fourier harmonics but with the coefficients squared.

Here we extend our approach~\cite{lokhtin04}, considering the energy-energy 
correlator
in the restricted rapidity-angle space in the vicinity of maximum energy
deposition of every event (Fig. 1). 
\begin{figure}
\includegraphics[width=80mm]{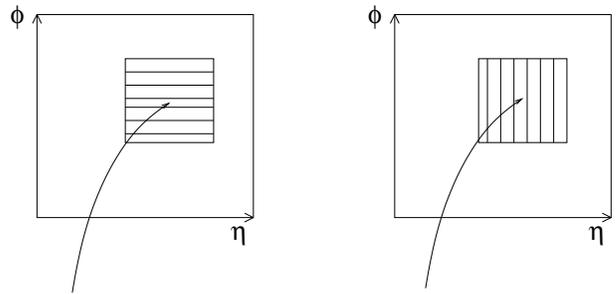}
\caption{The region of $\eta\times\phi$ space used in order to calculate
correlators in the near-side jet region.}
\end{figure}
It is convenient to define this special
correlator in the following form:
\begin{eqnarray}
\label{sigma-eta}
\frac{d\Sigma_T(\eta)}{d\eta}=&&\frac{1}{\Delta\eta~ N_{\rm
event}}\sum\limits_{{\rm event}}\sum\limits_{kl} \frac{E_{Tk}
E_{Tl}}{(E_T^{\rm vis})^2-\sum\limits_jE_{Tj}^2}\nonumber\\
&&\times\delta(k-l-m),
\end{eqnarray}
where $E_{Tk}$ is the transverse energy deposition in the pseudorapidity strip $k$
($k=-M,...,-1,0,+1,...,+M$) with the width $\Delta\eta \sim 0.1$ and the position
of a jet axis in the strip
$k=0$, summarized over all azimuthal sectors
$i$  ($i=-M,...,-1,0,+1,...,+M$) with the size $\Delta \phi = \Delta \eta$
and the position of a jet axis in the angle sector $i=0$;
$m=[\eta/\Delta\eta]$ (the integer part of the number $\eta/\Delta\eta$). Here 
$E_T^{\rm vis}=\sum\limits_lE_{Tl}$ is the total 
transverse energy deposition in the square
of $\eta\times\phi$ space, the center of which is determined by the position of
a jet axis. The similar definition can be written for the correlator as a
function of the azimuthal angle ( it is necessary to replace
$\eta \leftrightarrow \phi$ in Eq. (\ref{sigma-eta}) only).  The autocorrelation
term $\sum\limits_lE_{Tl}^2$ is subtracted in the denominator of the definition
in order to normalize the  integral (the sum) of correlator over  $\eta$  to 
one in the fixed square around the jet with the number of pseudorapidity strips
and angle sectors $(2M+1)$.

\section{A MODEL FOR ENERGY LOSS OF A JET}

We demonstrate the productivity and effectiveness of such correlators in the
framework of our well worked-out model of jet passing through a medium. The
model has been early
applied  to the calculation of various observables sensitive to the 
partonic energy loss: the impact parameter dependence of the jet production
~\cite{lokhtin00}, the mono/dijet rate enhancement~\cite{lokhtin99, bass},
the dijet rate dependence on the angular jet cone~\cite{lokhtin98}, 
the elliptic coefficient of the jet azimuthal anisotropy~\cite{lokhtin02, note, Lokhtin02, Lokhtin03}, 
the anti-correlation between  the softening jet fragmentation
function and the suppression of the jet rate~\cite{lokhtin03}. The validity of
approximations used has been verified by the comparison of RHIC data with
the model calculations. The agreement is good~\cite{lokhtin05} at the reasonable choice of main
model parameters taking into account their dependence on the collision
energy. At present the model has been constructed  as the fast Monte Carlo event
generator PYQUEN (PYthia QUENched), and corresponding Fortran routine pyquen is
available via Internet~\cite{pyquen}. The routine is implemented as a
modification of the  standard PYTHIA jet events~\cite{pythia}.
For details one can
refer to these mentioned papers (mainly,~~\cite{lokhtin00,lokhtin98,lokhtin03,
lokhtin05}).
Here we note only the main steps essential for the present investigation. 

The energy-energy correlator depends on not only the absolute value of partonic
energy loss, but also on the angular spectrum of in-medium radiated gluons. 
Since coherent Landau-Pomeranchuk-Migdal radiation induces a strong dependence 
of the radiative energy loss of a jet on the angular cone 
size~\cite{lokhtin98,baier,Zakharov:1999,urs,vitev}, it will soften particle 
energy distributions inside the jet, increase the multiplicity of secondary 
particles, and to a lesser degree, affect the total jet energy. On the other 
hand, collisional 
energy loss turns out to be practically independent of the jet cone size 
and causes 
the loss of total jet energy, because the bulk of ``thermal'' particles 
knocked out of the 
dense matter by elastic scatterings fly away in an almost 
transverse direction relative to 
the jet axis~\cite{lokhtin98}. Thus although the radiative energy loss of an 
energetic parton dominates over the collisional loss by up to an order of 
magnitude, the relative contribution of the collisional loss of 
a jet grows with 
increasing jet cone size due to the essentially different angular 
structure of loss 
for two mechanisms~\cite{lokhtin98}. Moreover, the total energy loss of a jet 
will be sensitive to the experimental capabilities to detect low-p$_T$ 
particles -- products of soft gluon fragmentation: thresholds for a giving signal 
in calorimeters, influence of the  strong magnetic field, etc.~\cite{baur}. 

Since the full treatment of the angular spectrum of emitted gluons is rather 
sophisticated and 
model-dependent~\cite{lokhtin98,baier,Zakharov:1999,urs,vitev}, we considered 
two simple parameterizations of the  distribution of in-medium radiated gluons over
the emission angle $\theta$. The ``small-angular'' radiation spectrum was
parameterized in the form
\begin{equation} 
\label{sar} 
\frac{dN^g}{d\theta}\propto \sin{\theta} \exp{\left( -\frac{(\theta-\theta
_0)^2}{2\theta_0^2}\right) }~, 
\end{equation}
where $\theta_0 \sim 5^0$ is the typical angle of the coherent gluon radiation estimated
in Ref.~\cite{lokhtin98}. The ``broad-angular'' spectrum has the form 
\begin{equation} 
\label{war} 
\frac{dN^g}{d\theta}\propto \frac{1}{\theta}~.  
\end{equation}
We believe that such a simplified treatment here is enough to demonstrate
the sensitivity of the energy-energy correlator to
the medium-induced partonic energy loss. 

PYTHIA$\_6.2$~\cite{pythia} was used to generate the initial jet distributions 
in nucleon-nucleon sub-collisions at $\sqrt{s}=5.5$ TeV.
After that, event-by-event Monte Carlo simulation of rescattering and energy 
loss of jet partons in QGP was performed. The approach relies on   
accumulative energy losses, when gluon radiation is associated with each 
scattering in the expanding medium together with including the interference effect by 
the modified radiation spectrum as a function of decreasing temperature.
Such a numerical simulation of the free path of a hard jet in QGP allows any 
kinematical characteristic distributions of jets in the final state to be 
obtained. Besides, the different scenarios of medium evolution can be considered. 
In each $i$th scattering a fast parton loses energy 
collisionally and radiatively, $\Delta e_i = t_i/(2m_0) + \omega _i$, 
where the transfer momentum squared $t_i$ is simulated according to the
differential cross section for elastic
scattering of a parton with energy $E$ off the 
``thermal'' partons with energy (or effective mass) $m_0 \sim 3T \ll E$ 
at temperature $T$, and
$\omega _i$ is simulated according to the energy spectrum of coherent
medium-induced gluon radiation in the Baier-Dokshitzer-Mueller-Schiff
formalism~\cite{baier}. 
Finally we suppose that in every event the energy of an initial parton 
decreases by the value $\Delta E= \sum _i \Delta e_i$. 

The medium was treated as a boost-invariant longitudinally expanding 
quark-gluon fluid, 
and partons as being produced on a hyper-surface of equal proper times 
$\tau$~\cite{bjorken}. For certainty we used the initial conditions 
for the gluon-dominated plasma formation 
expected for central Pb$+$Pb collisions at LHC~\cite{esk}: 
$\tau_0 \simeq 0.1$ fm/$c$, $T_0 \simeq 1$ GeV. For non-central collisions we 
suggest the proportionality of the initial energy density to the ratio of 
the nuclear overlap function 
and the effective transverse area of nuclear overlapping~\cite{lokhtin00}.

In the frame of this model and using above QGP parameters we evaluate the mean
energy loss of quark of $E_T=100$ GeV in minimum-bias Pb$+$Pb collisions,
$<\Delta E^q_T> \sim 25$ GeV (accordingly, the gluon energy loss is 
$<\Delta E^g_T>=9/4<\Delta E^q_T>$). In our simulation the value of energy loss
in fact is regulated mainly by the initial temperature $T_0$.

\section{NUMERICAL RESULTS AND DISCUSSION}

To be specific, we consider the geometry of the Compact Muon Solenoid (CMS)
detector~\cite{baur} at LHC. The central (barrel) part of the CMS 
calorimetric system  covers the pseudorapidity region $|\eta| < 1.5$, the 
segmentation of electromagnetic and hadron calorimeters being $\Delta \eta 
\times \Delta \phi = 0.0174 \times 0.0174$ and $\Delta \eta \times \Delta \phi 
= 0.0872 \times 0.0872$ respectively~\cite{baur}. The endcap part of the CMS
calorimeter $1.5 < |\eta| < 3$ has more tangled structure: the tower sizes of
hadron calorimeter over $\eta$ and $\phi$ are 0.087 at $|\eta| = 1.5$ and
increase gradually up to 0.345 at $|\eta| = 3$.

Then the energy-energy correlation function 
(\ref{sigma-eta}) is calculated for the events containing at least one jet with 
$E_T^{\rm jet} > E_T(\rm threshold) = 100$ GeV in the considered kinematical 
region. The final jet energy is defined here as the total transverse energy of 
final particles collected around the direction of a leading particle inside the 
cone $R=\sqrt{\Delta \eta ^2+\Delta \phi ^2}=0.7$, where $\eta$ and 
$\phi$ are the pseudorapidity and the azimuthal angle respectively. Note
that the estimated event rate in the CMS acceptance, $\sim 10^7$ jets with
$E_T>100$ GeV in a one month LHC run with lead beams~\cite{Accardi:2003,baur}, will be 
large enough to carefully study the energy-energy correlations over the whole
azimuthal range.  

We restrict
ourself to central collisions in which the effect of jet quenching is maximum
and take into consideration the energy deposition resulting from jets only. 
The fact is that in the CMS heavy ion physics program the modified sliding
window-type jet finding algorithm has been developed to search for ``jet-like''
clusters above the average energy and to subtract the background from the
underlying event~\cite{Accardi:2003,baur, vardanian05}. This developed algorithm
allows one to reconstruct the jet energy and its space position with a high enough
accuracy in a high multiplicity environment beginning from some threshold jet
transverse energy $E_T^{\rm jet} \sim 50-100$ GeV. Thus, for instance, the
reconstruction efficiency (the ratio of the number of reconstructed jets to the
total number of generated jets) for jets with energy $E_T^{\rm jet} > 100$ GeV
is close to $100$\%, the fraction of false jets being not greater than $1$\% 
of the total number of reconstructed jets. The mean reconstructed jet energy
$E^{\rm jet}_T({\rm reco})$ is a linear function of the mean generated jet
energy $E^{\rm jet}_T({\rm MC})$, the shape of this dependence being identical
in the presence and in the absence of the background from Pb$+$Pb events. This
means that $pp$-collisions can be used for an adequate comparison with Pb$+$Pb
collisions. The attainable precision of the space resolution of jets is rather
high: it is less than the size of a hadron calorimeter 
tower~\cite{Accardi:2003,baur, vardanian05}. 

This gives grounds to hope that in the region of maximum energy deposition the
contributions from jets and ``thermal'' background can be effectively separated
not only on the simulation level, when they are known {\it a priori}. Therefore in
order to reveal the new possible issues of the investigation of energy correlations we
neglect this possible difference between the initially generated and
reconstructed jet energy depositions, which can be significant on the outlying jet
area only. Though the analysis of the  quality of extracting the jet-like 
energy-energy correlator against background fluctuations based on the detailed 
simulation of detector responses (together with the optimization of the jet 
detection threshold and the calibration coefficients used at the subtraction and
reconstruction procedures) should be performed for each specific experiment.  
In particular, in CMS we have to take into consideration that most of the charged 
low-$p_T$ particles will be cleared out of the central calorimeters by the 
strong magnetic field (decreasing a really measurable energy flux 
and absolute value of its fluctuations in such a way).

The results of the numerical simulations are presented in Fig.~2 (for the cases 
without and with partonic energy loss) for the events containing at least one
hard jet with  $E_T^{\rm jet} > E_T(\rm threshold) = 100$ GeV.
\begin{figure}
\includegraphics[width=90mm]{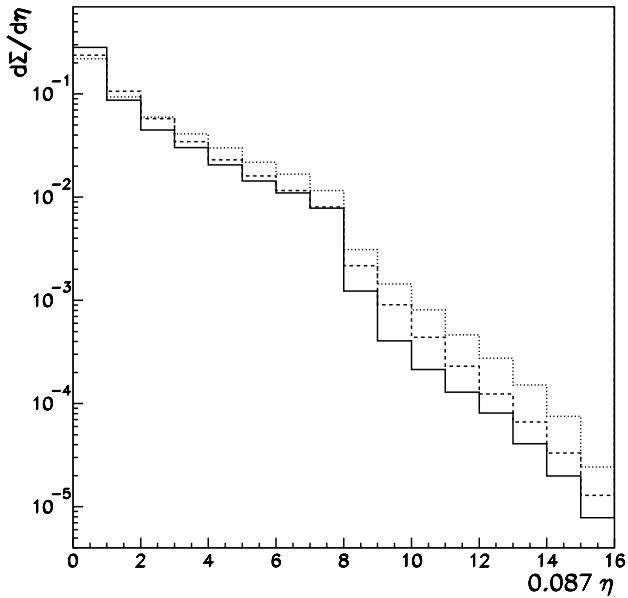}
\caption{The transverse energy-energy correlator $d\Sigma_T/d\eta$ 
in the near-side jet region as a function of pseudorapidity $\eta$
without (solid histogram) and with medium-induced partonic 
energy loss for the ``small-angular'' (\ref{sar}) (dashed histogram) and the 
``broad-angular'' (\ref{war}) (dotted histogram) parameterizations of emitted 
gluon spectrum in central Pb$+$Pb collisions for a jet axis position from the
central pseudorapidity region $|\eta^{\rm jet}|<0.8$.}
\end{figure}
The two parameterizations of the distribution on the gluon emission 
angles (\ref{sar}) and (\ref{war}) were used and                                            
$\Delta\eta=\Delta\phi=\Delta\varphi=0.087, M=8$. One can see that the energy
loss results in the noticeable broadening of jet shape which can be
characterized by the variation of the mean pseudorapidity squared (the width of
distribution). These mean values of pseudorapidity squared are defined in the
following form:
\begin{equation}
\label{eta2}
<\eta^2>~=~ \int d\eta~ \eta^2~ \frac{d\Sigma_T(\eta)}{d\eta}\Biggl/
\int d\eta~ \frac{d\Sigma_T(\eta)}{d\eta}
\end{equation}
and are displayed in Table I.
They are mainly determined by the first bins of histograms closed to a jet axis  
where the separation of jet and background contributions to the total energy
deposition is expected to be effective. The broadening effect is considerable
for the ``broad-angular'' radiation,  comes to  $\sim 45$\% and probably can be
observable in spite of the ambiguities discussed above.
\begin{table*}
\caption{\label{tab:table1}The mean values of pseudorapidity and angle squared as a measure
of the width of transverse energy-energy correlation functions.}
\begin{ruledtabular}
\begin{tabular}{c c c c c c}   
& & Energy loss (\ref{sar})  & Energy loss (\ref{war}) 
& Energy loss(\ref{sar})  & Energy loss (\ref{war})  \\   
Model & Without loss & off comoving & off comoving & off ``slow'' & 
off ``slow'' \\ 
& & constituents & constituents & constituents & constituents \\ \hline
\multicolumn{6}{c}{$|\eta^{\rm jet}| < 0.8$} \\  
$<\eta^2>$ & 0.029 & 0.033 & 0.042 & 0.034 & 0.051 \\  
$<\varphi^2>$ & 0.030 & 0.034 & 0.044 & 0.035 & 0.054 \\ \hline 
\multicolumn{6}{c}{$0.8 < \eta^{\rm jet} < 2.4$} \\ 
$<\eta^2>$ & 0.028 & 0.032 & 0.044 & 0.046 & 0.055 \\ 
$<\varphi^2>$ & 0.028 & 0.033 & 0.045 & 0.048 & 0.056 \\ 
\end{tabular}
\end{ruledtabular}
\end{table*}

Figure 3 shows the behavior of correlator (\ref{sigma-eta}) in the near-side
jet region as a function of azimuthal angle $\varphi$ (the notations and the
kinematical restrictions are the same as in Fig. 2). 
\begin{figure}
\includegraphics[width=90mm]{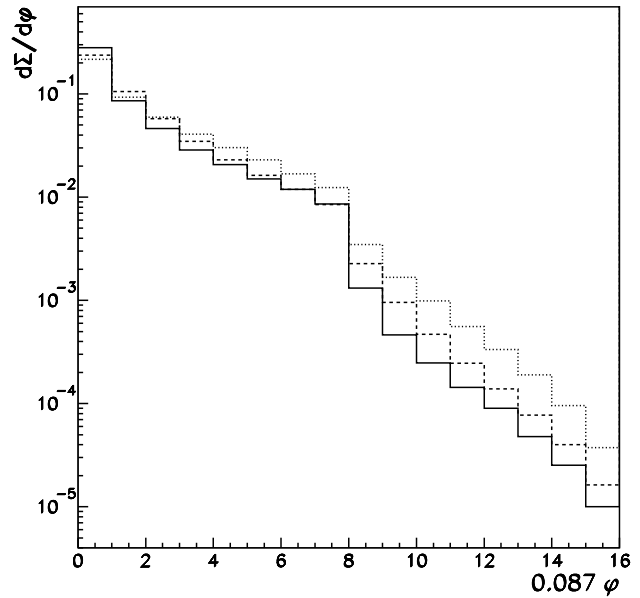}
\caption{The transverse energy-energy correlator $d\Sigma_T/d\varphi$ 
in the near-side jet region as a function of azimuthal angle $\varphi$
without (solid histogram) and with medium-induced partonic 
energy loss for the ``small-angular'' (\ref{sar}) (dashed histogram) and the 
``broad-angular'' (\ref{war}) (dotted histogram) parameterizations of emitted 
gluon spectrum in central Pb$+$Pb collisions for a jet axis position from the
central pseudorapidity region $|\eta^{\rm jet}|<0.8$.}
\end{figure}
As in the previous case the
broadening is more pronounced for the ``broad-angular'' parameterization of
emitted gluon spectrum. Besides the equality (with a good accuracy) of mean
values of pseudorapidity and azimuthal angle squared results from 
the rapidity-angle symmetry of
jet shape for jets from the central pseudorapidity region. For the illustration the
difference between correlators 
$$\Bigl(\frac{d\Sigma_T(\eta)}{d\eta}- \frac{d\Sigma_T(\varphi)}{d\varphi}
\Bigr)\Big|_{\eta=\varphi}$$ as a function of $\eta=\varphi$ is presented in
Fig. 4. This difference is small, nonregular and comes to the level of
statistical fluctuations.
\begin{figure}
\includegraphics[width=90mm]{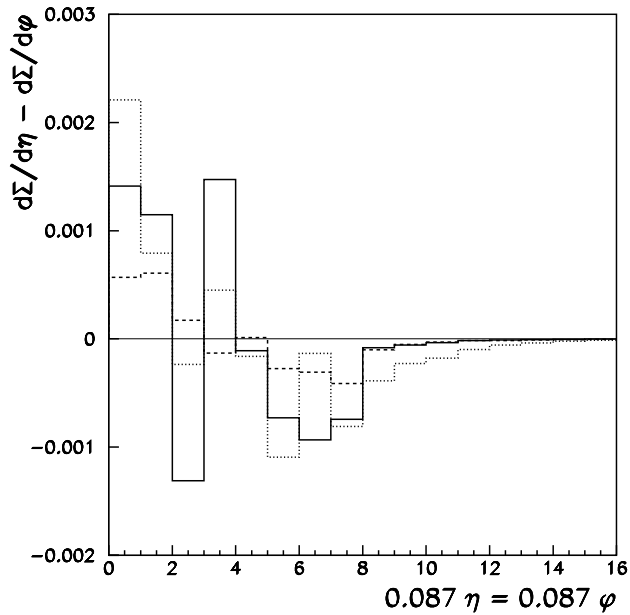}
\caption{The 
difference between correlators 
$\Bigl(\frac{d\Sigma_T(\eta)}{d\eta}- \frac{d\Sigma_T(\varphi)}{d\varphi}
\Bigr)\Big|_{\eta=\varphi}$ 
in the near-side jet region as a function of $\eta=\varphi$
without (solid histogram) and with medium-induced partonic 
energy loss for the ``small-angular'' (\ref{sar}) (dashed histogram) and the 
``broad-angular'' (\ref{war}) (dotted histogram) parameterizations of emitted 
gluon spectrum in central Pb$+$Pb collisions for a jet axis position from the
central pseudorapidity region $|\eta^{\rm jet}|<0.8$.}
\end{figure}

We are interested in the nonsymmetrical behavior of correlators as functions of
pseudorapidity and azimuthal angle because there is some discussion in the current
literature~\cite{urs04} about the possible nonsymmetrical modification of jet
shape due to the collective motion of medium constituents, on which the
rescattering of jet partons takes place. The correlators under consideration
could show the different behavior over angle and pseudorapidity for a jet axis
position from the nonsymmetrical pseudorapidity region. However our simulation gives
practically the former results for correlators calculated 
for jets from the noncentral
pseudorapidity region $0.8<\eta^{\rm jet}<2.4$. The corresponding values of
pseudorapidity
and angle squared are listed  in the lower part of Table I. The
difference between correlators as a function of $\eta=\varphi$ is again small,
nonregular and comes to the level of statistical fluctuations as in Fig.~4.

Thus the behavior of transverse energy-energy correlators in the near-side jet
region is independent of the pseudorapidity position of a jet axis. The fact is that
in our model the rescattering of a hard parton takes place off a comoving  medium
constituent (i.e. moving with the same longitudinal rapidity as a hard parton).
In such approach gluons are emitted at the polar angle $\theta$ relative to the
{\it transverse
momentum} $p_T$ of a radiating parton isotropically over the azimuthal angle in
the rest frame of a medium constituent that, as a matter of fact, provides for
the longitudinal boost  invariance of correlators. However this scenario with
the same longitudinal rapidity of a jet parton and a medium constituent (mostly
used and closely related with the Bjorken scaling solution) is not unique and
one can suppose that a jet parton and a medium constituent move in the
longitudinal direction in a different way.

On the generator level it is not difficult to realize the scenario when the
rescattering takes place off a so-called ``slow'' medium constituent (i.e. moving with
the zero or considerably less longitudinal rapidity in comparison with the rapidity
of a jet parton) and gluons are emitted at the polar angle $\theta$ relative to the
{\it total momentum} $p$ of a parent parton.
From Table I one can see that for jets from the central pseudorapidity region 
$|\eta^{\rm jet}|<0.8$ and the ``small-angular'' parameterization (\ref{sar})
this modification does not lead to any  variation of correlation
functions because of the small distinction between the total 
and transverse momenta in this case.
For the ``broad-angular'' radiation (\ref{war}) the broadening of jet shape
is independent of the rescattering scenario in the central pseudorapidity region at
the additional restriction on the difference between the directions of an initial
hard parton and a final jet axis (due to the large ``tail'' in the distribution over this
difference). 

For jets from the forward pseudorapidity region 
$0.8<\eta^{\rm jet}<2.4$ (where the distinction between the total and 
transverse momenta is significant) the broadening of jet shape increases
considerably that is supported by the calculated mean values of pseudorapidity and 
angle squared. This larger broadening results from that the energy loss is 
determined by not the transverse momentum (as this was early) 
but the total one which is
considerably larger than the transverse momentum for jets from the forward
pseudorapidity 
region. The experimental observation of such dependence of the width of
correlation functions 
on the pseudorapidity position of a jet axis will indicate the scenario with the
gluon radiation around the total momentum of a parent parton. Besides this
observation does not demand the comparison with  $pp$-collisions, in which
such dependence is absent. 

As regards the possible $\eta\times\varphi$ asymmetry,
then it is invisible as before on the level of the mean values, although we
might catch some regular (but  again practically on the level of statistical 
fluctuations) difference in the behavior of correlators over $\eta$
and $\varphi$. This can mean that the transverse energy-energy correlations
(unlike the average jet energy and jet multiplicity~\cite{urs04}) is
weakly sensitive to longitudinal flow effects due to the longitudinal Lorentz
invariance of transverse energy itself or our simple imitation of flow effects 
on the generator level can be, in principle, not enough to their study  in respect of
the possible $\eta\times\varphi$ asymmetry.

The scenarios suggested above result also in the different dependence of jet
quenching on the pseudorapidity. In the accepted scenario (I) with the
rescattering off comoving medium constituents jet quenching is expected to be
independent of the jet pseudorapidity under the condition that the matter density is
also $\eta$-independent
\begin{figure}
\includegraphics[width=90mm]{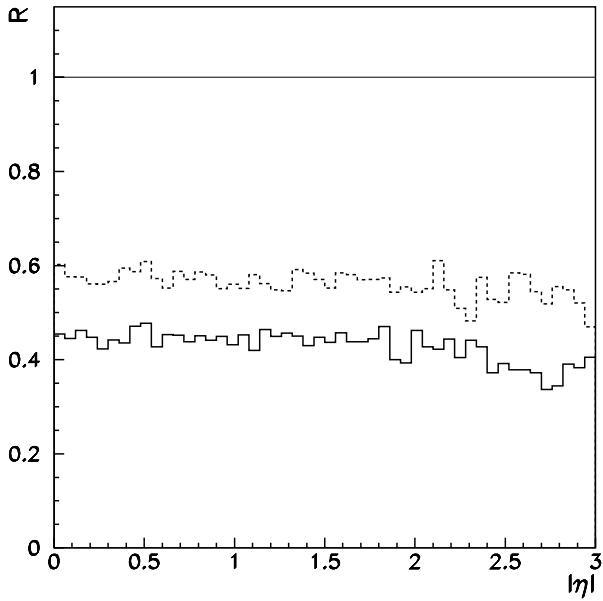}
\caption{The jet quenching factor $R$ 
as a function of pseudorapidity $\eta$
without (solid line) and with medium-induced partonic 
energy loss for the ``small-angular'' (\ref{sar}) (dashed histogram) and the 
``broad-angular'' (\ref{war}) (solid histogram) parameterizations of emitted 
gluon spectrum in central Pb$+$Pb collisions in scenario I.}
\end{figure}
and Figure 5 demonstrates this independence.
We define a jet quenching factor $R$ in a conventional way,
\begin{equation}
\label{R}
R =  \frac{N(E_{\rm jet} > 100~{\rm GeV})|_{ {\rm with~losses}}}
{N(E_{\rm jet} > 100~{\rm GeV})|_{ {\rm without~losses}}},
\end{equation}
as the ratio of the jet number with transverse energy $E_T^{\rm jet} > 100$ GeV
with energy loss to the corresponding number of jets without loss 
(i.e. in $pp$-collisions normalized on the number of binary nucleon-nucleon
sub-collisions). In the second scenario (II) a quenching factor is strongly (Fig.~6)
\begin{figure}
\includegraphics[width=90mm]{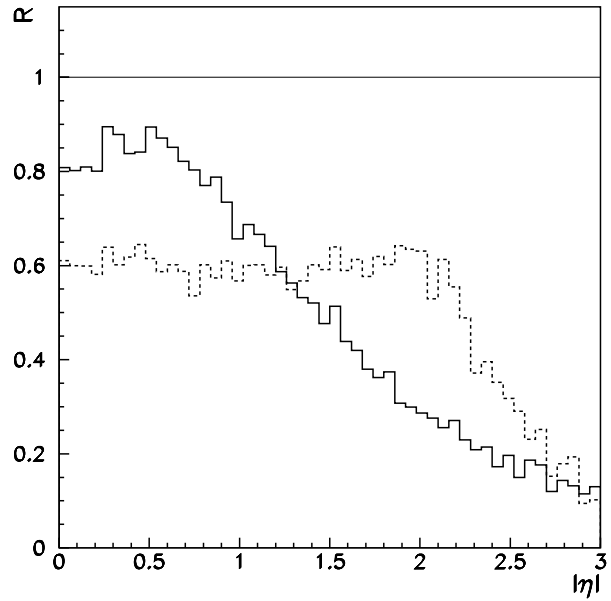}
\caption{The jet quenching factor $R$ 
as a function of pseudorapidity $\eta$
without (solid line) and with medium-induced partonic 
energy loss for the ``small-angular'' (\ref{sar}) (dashed histogram) and the 
``broad-angular'' (\ref{war}) (solid histogram) parameterizations of emitted 
gluon spectrum in central Pb$+$Pb collisions in scenario II.}
\end{figure}
dependent on the pseudorapidity position of a jet axis. For the ``broad-angular''
radiation this factor $R$ has more tangled behavior with the shifted (not in
the zero) maximum over $\eta$. This can be explained by the large ``tail'' in
the distribution over the difference (especially in $\eta\times\varphi$ space !)
between the directions of an initial hard 
parton and a final jet axis and by our simplified jet definition. Here  
the final jet energy is defined  simply as the total transverse energy of 
final particles collected around the direction of a leading particle inside the 
cone $R=\sqrt{\Delta \eta ^2+\Delta \phi ^2}=0.7$ without correction of the jet
axis usually used in the modified sliding window-type jet finding algorithm.
The structure with the pseudorapidity-shifted maximum disappears
if the restriction on the difference between the directions of an initial hard 
parton and a final jet axis is established as one can see in Fig.~7 or if
\begin{figure}
\includegraphics[width=90mm]{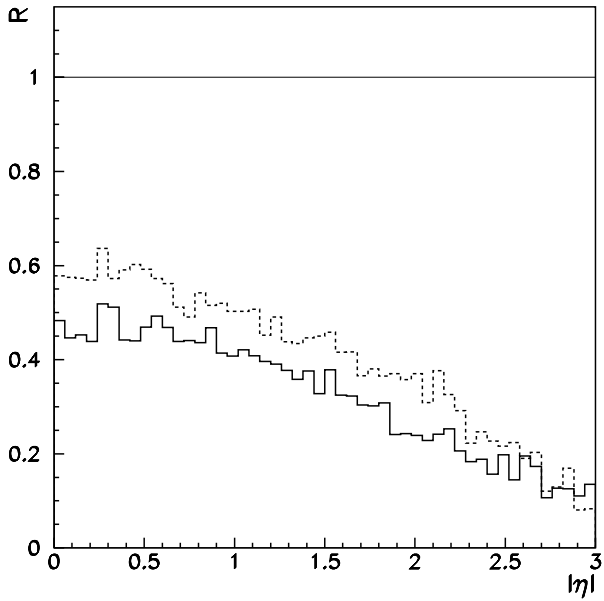}
\caption{The jet quenching factor $R$ 
as a function of pseudorapidity $\eta$
without (solid line) and with medium-induced partonic 
energy loss for the ``small-angular'' (\ref{sar}) (dashed histogram) and the 
``broad-angular'' (\ref{war}) (solid histogram) parameterizations of emitted 
gluon spectrum in central Pb$+$Pb collisions in scenario II under the additional
restriction on the difference between the directions of an initial hard parton
and a final jet axis ($<0.2$).}
\end{figure}
the modified sliding
window-type jet finding algorithm~\cite{Accardi:2003,baur, vardanian05} is used.
But for all that the additional restriction and the use of other algorithm
have no influence on the rapidity-behavior of
quenching factor in the scenario I. Thus the quenching factor together with the
jet shape broadening is independent of the pseudorapidity position of a jet axis in
the scenario I while in the scenario II the dependence of these observables on 
the pseudorapidity is noticeable and probably can be experimentally detected in spite 
of ambiguities discussed above.

\section{SUMMARY}
 
The special correlator in the vicinity of maximum energy deposition of every
event allowed us to investigate the jet shape modification due to partonic
energy loss using the calorimetric information. 
In the accepted scenario with scattering of jet hard partons off  
comoving medium constituents this correlator is independent of the pseudorapidity 
position of a jet axis and becomes considerably broader (symmetrically over the 
pseudorapidity and the azimuthal angle) in comparison with $pp$-collisions.
At scattering off ``slow''  medium constituents the broadening of 
correlation functions is dependent on the pseudorapidity position of a jet axis 
and increases  noticeably in comparison with the previous scenario for jets with 
large enough pseudorapidities. These two considered scenarios result also in the
different dependence of jet quenching on the pseudorapidity.
We believe that such a transverse  energy-energy correlation analysis may be
useful at LHC data processing.

\begin{acknowledgments}
Discussions with D.~Lopez, S.V.~Molodtsov, C.~Roland, I.N.~Vardanyan, 
B.~Wyslouch, and G.M.~Zinovjev are gratefully
acknowledged. This work is supported 
partly by Grant INTAS-CERN 05-112-5475 and by
Russian Ministry of Science and Education, Contract N 02.434.11.7074.
\end{acknowledgments}


\end{document}